\documentclass[10pt,preprint,aps,prb,showpacs,superscriptaddress]{revtex4-1}
\usepackage{epsfig}
\usepackage{float}
\usepackage{color}
\usepackage{amssymb}
\begin{document}

\title{Fermi surface reconstruction in (Ba$_{1-x}$K$_x$)Fe$_2$As$_2$ (0.44 $\leq x \leq$ 1) probed by thermoelectric power measurements}

\author{Halyna Hodovanets}
\affiliation{Department of Physics \& Astronomy, Iowa State University, Ames, IA 50011, USA}
\affiliation{Division of Materials Sciences and Engineering, Ames Laboratory, Ames, Iowa 50011, USA}

\author{Yong Liu}
\affiliation{Division of Materials Sciences and Engineering, Ames Laboratory, Ames, Iowa 50011, USA}

\author{Anton Jesche}
\affiliation{Department of Physics \& Astronomy, Iowa State University, Ames, IA 50011, USA}
\affiliation{Division of Materials Sciences and Engineering, Ames Laboratory, Ames, Iowa 50011, USA}

\author{Sheng Ran}
\affiliation{Department of Physics \& Astronomy, Iowa State University, Ames, IA 50011, USA}
\affiliation{Division of Materials Sciences and Engineering, Ames Laboratory, Ames, Iowa 50011, USA}

\author{Eun Deok Mun}
\affiliation{Department of Physics \& Astronomy, Iowa State University, Ames, IA 50011, USA}

\author{Thomas A. Lograsso}
\affiliation{Division of Materials Sciences and Engineering, Ames Laboratory, Ames, Iowa 50011, USA}
\affiliation{Department of Materials Science and Engineering, Iowa State University, Ames, Iowa 50011, USA}

\author{Sergey L. Bud'ko}
\affiliation{Department of Physics \& Astronomy, Iowa State University, Ames, IA 50011, USA}
\affiliation{Division of Materials Sciences and Engineering, Ames Laboratory, Ames, Iowa 50011, USA}

\author{Paul C. Canfield}
\affiliation{Department of Physics \& Astronomy, Iowa State University, Ames, IA 50011, USA}
\affiliation{Division of Materials Sciences and Engineering, Ames Laboratory, Ames, Iowa 50011, USA}

\begin{abstract}

We report in-plane thermoelectric power measurements on single crystals of (Ba$_{1-x}$K$_x$)Fe$_2$As$_2$ (0.44 $\leq x \leq$ 1). We observe a minimum in the $S\mid _{T=const}$ versus $x$ at $x \sim$ 0.55 that can be associated with the change in the topology of the Fermi surface, a Lifshitz transition, related to the electron pockets at the center of $M$ point crossing the Fermi level. This feature is clearly observable below $\sim$ 75 K. Thermoelectric power also shows a change in the $x \sim$ 0.8 - 0.9 range, where maximum in the thermoelectric power collapses into a plateau. This Lifshitz transition is most likely related to the reconstruction of the Fermi surface associated with the transformation of the hole pockets at the $M$ point into four blades as observed by ARPES measurements.  
\end{abstract}


\maketitle

\section{Introduction}

Since their discovery\cite{Kamihara2008}, iron-based superconductors have been the subject of extensive theoretical and experimental research in order to better understand their physical properties and, more importantly, the origin of superconductivity. AEFe$_2$As$_2$ (AE - alkali earth) is one of the most studied structural families in part because it allows for the study of AE, Fe, and As site substitutions.\cite{Canfield2010a,Mandrus2010,Ni2011} Among various members, (Ba$_{1-x}$K$_x$)Fe$_2$As$_2$ has the highest bulk $T_c$ of $\approx$ 38 K\cite{Rotter2008} and doping is done in the Ba-layers with the least perturbation to the FeAs layers, which are believed to be essential for superconductivity. Interestingly, in this series, the superconducting dome extends to, and includes, the end member, KFe$_2$As$_2$ with $T_c\approx$ 3.5 K.\cite{Avci2012}

Learning about the Fermi surface (FS) is important for understanding of the mechanism of superconductivity.\cite{Mazin2009,Kemper2010,Hirschfeld2011,Chubukov2012}  Based on the angle-resolved photoemission spectroscopy (ARPES) studies of (Ba$_{1-x}$K$_x$)Fe$_2$As$_2$,\cite{Liu2008,Sato2009, Yoshida2011,Xu2011,Nakayama2011,Richard2011,Okazaki2012,Malaeb2012,Xu2013,Ota2014}  one or two Lifshitz transitions\cite{Lifshitz1960} are expected between $x \sim0.4$ and  $x$= 1. One Lifshitz transition is associated with the disappearance of the electron pockets around the zone corner ($M$ point)\cite{Malaeb2012} and second Lifshitz transition is associated with the transformation of the ellipsoid-like hole pockets at the $M$ point into small off-$M$ centered hole FS pocket lobes\cite{Xu2013}. Both Ref. \onlinecite{Malaeb2012} and Ref. \onlinecite{Xu2013} report a Lifshitz transition, albeit at different K-concentrations. Ref. \onlinecite{Malaeb2012} claims the disappearance of the electron pocket at $M$ point. Ref. \onlinecite{Nakayama2011} states that the shallow electron pocket is still present at $M$ point for $x$ = 0.7 and is expected to cross the Fermi energy upon further K-substitution in the rigid band approximation. Ref. \onlinecite{Xu2013} claims the emergence of four small off-$M$-centered lobes with some, within error bars, intensity due to electron pocket at the $M$ point. These results seem contradictory and inconclusive.

Interestingly, an anomaly in the pressure derivatives and deviation from the $\Delta C_p/T_c\sim T_c^3$, known as BNC scaling,\cite{Budko2009} were observed at $x \sim$ 0.7 by pressure dependent magnetization and specific heat measurements.\cite{Budko2013,Stewart2011} Moreover, an  electronic topological (Lifshitz) transition was predicted at $x \sim$ 0.9 for (Ba$_{1-x}$K$_x$)Fe$_2$As$_2$ by the band structure calculations.\cite{Khan2014} And a change in the superconducting gap symmetry was observed at $x \sim 0.76$ by thermal conductivity and penetration depth measurements.\cite{Watanabe2014}

Thermoelectric power (TEP) can be very sensitive to the changes in the FS because, grossly speaking, TEP depends, in part, on the derivative of the density of states at the Fermi level. Hence, TEP measurements can be a good tool for probing the potential changes in the FS topology of (Ba$_{1-x}$K$_x$)Fe$_2$As$_2$. TEP was measured on polycrystalline samples of hole-doped (Ba,K)Fe$_2$As$_2$ \cite{Yan2010} and it was concluded that $S$(300 K) as a function of K-concentration has a complex behavior and may manifest features consistent with the existence of strong spin fluctuations in heavy-doped K-samples. Although transport measurements on polycrystalline samples are useful, most of the time the anisotropic dependencies of the properties are averaged and subtle changes can be smeared and missed. Therefore, we performed TEP measurements on single crystals of (Ba$_{1-x}$K$_x$)Fe$_2$As$_2$ (0.44 $\leq x \leq$ 1) with $\bigtriangledown T\| \bf{ab}$ on a denser set of K-concentrations on the over-doped side of the phase diagram. 
We observe features at $x \sim$ 0.55 and $x \sim$ 0.8-0.9 that are consistent with the Lifshitz transitions that have been associated with the shift of the electron pockets at the $M$ point above the Fermi level and the transformation of the hole pockets near the $M$ point into "four blades" as observed by ARPES measurements\cite{Xu2013,Malaeb2012}.

\section{Experimental Details}

(Ba$_{1-x}$K$_x$)Fe$_2$As$_2$ single crystals were grown following the procedure outlined in detail in Ref. \onlinecite{Liu2014}. An $x_{nom} \sim$ 0.55 batch of (Ba$_{1-x}$K$_x$)Fe$_2$As$_2$ single crystals was grown using Fe and As rich melt with an initial elemental composition of Ba:K:Fe:As = $1-x$:3$x$:4:5. A K- and As-rich melt was used to grow heavily K-doped (Ba$_{1-x}$K$_x$)Fe$_2$As$_2$ ($x_{nom}$ = 0.65, 0.80, 0.82, 0.92, and 1) batches of single crystals with initial elemental compositions of Ba:K:Fe:As = $y$:5:2:6 ($y$ = 0.1, 0.2, and 0.3). The elements were loaded into an alumina crucible, and then sealed in a tantalum tube by arc welding. Different soaking temperatures of $T$ = 920, 1000, and 1050 $^0$C were used to grow heavily K-doped (Ba$_{1-x}$K$_x$)Fe$_2$As$_2$ single crystals. The soaking temperature of $T$ = 920 $^0$C works well for growing (Ba$_{1-x}$K$_x$)Fe$_2$As$_2$ ($x_{nom}$ = 0.92 and 1) single crystals. However, it was found that higher soaking temperatures of $T$ = 1000 and 1050 $^0$C and the temperature window 900 $^0$C $< T <$ 1050 $^0$C are useful to grow high quality crystals with a sharp superconducting transition for the samples with $x_{nom}$ = 0.65 and 0.80. 

\begin{figure}[tbh]
\centering
\includegraphics[width=0.75\linewidth]{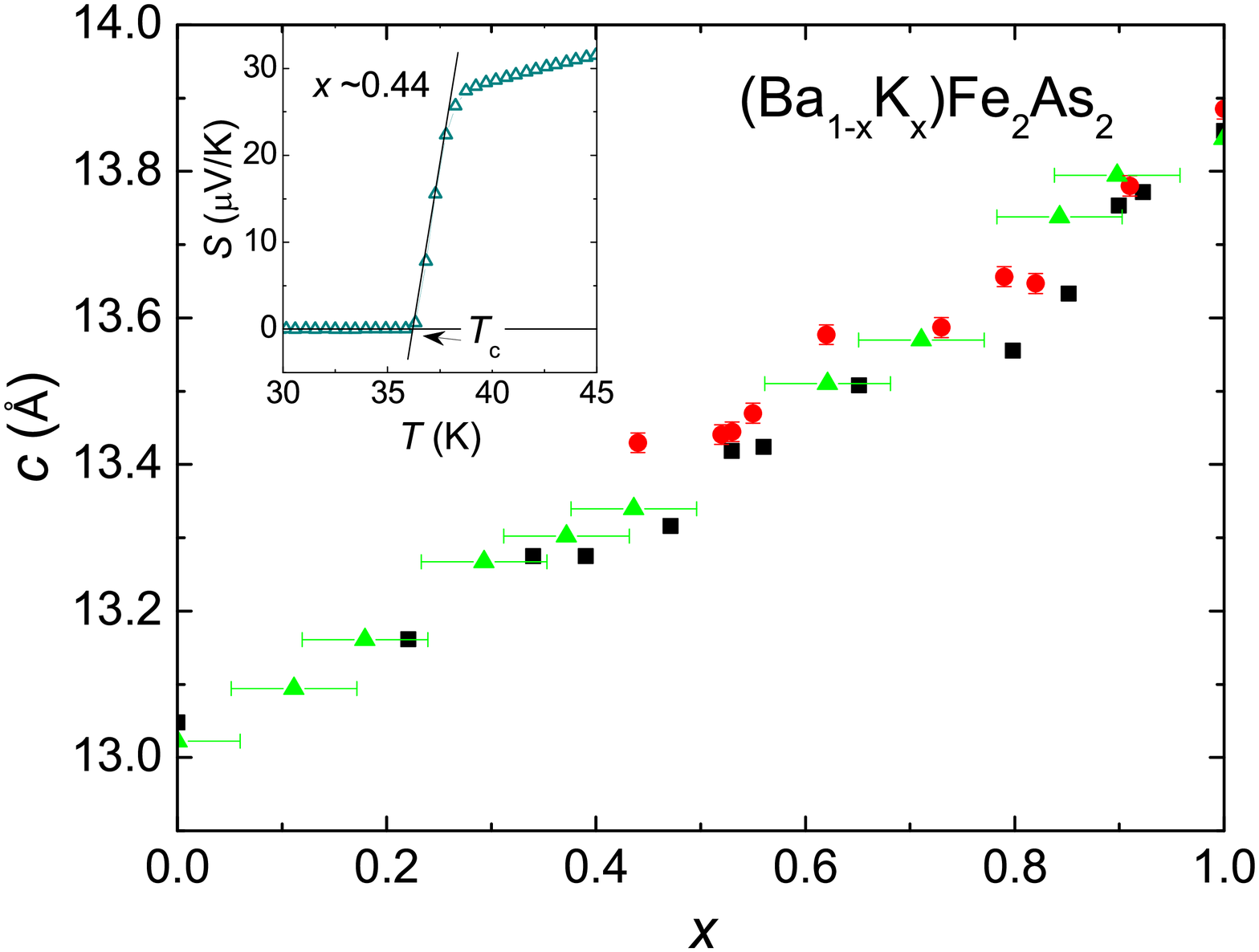}
\caption{\footnotesize (Color online) $c$-lattice parameter of (Ba$_{1-x}$K$_x$)Fe$_2$As$_2$ single crystals as a function of K-concentration. For comparison, the data from Refs. [\onlinecite{Liu2014}] (squares) and [\onlinecite{Rotter2008a}] (triangles) are also given. The data were collected on the same samples as were used for the thermoelectric power measurements. The inset shows an offset criterion used to determine $T_c$. The arrow denotes a $T_c$ for $x\sim$ 0.44.}
\label{c-all}
\end{figure}

From the earliest growth of (Ba$_{1-x}$K$_x$)Fe$_2$As$_2$ single crystals\cite{Ni2008b}, it was been appreciated that the degree of K-substitution can vary within the batch or even within the different layers of the same crystal. Indeed in this study, the single crystals picked from the same batch had different $T_c$ values. That is why the concentrations of potassium for the specific samples used in this paper were inferred by matching the $T_c$ values obtained from the offset criterion, shown in the inset to Fig. \ref{c-all}, in the thermoelectric power (TEP) measurements to that in the published phase diagram \cite{Avci2012}. Defining the K-concentration this way poses a difficulty in determining the error bars in the K-concentration. However, based on the sharpness of $T_c$, we expect the error bars associated with the value of $x$ determined this way to be smaller than 0.05.

The concentrations of K as determined by this $T_c$ criterion were also checked by measuring the $c$-lattice parameters of the same samples that where used for the TEP measurements. After the sample was dismounted from the TEP measurement set-up, the ends containing silver paste were carefully cut off and the tiny sample ($\sim$ 1-2 mm long and $\sim$ 0.1 mm wide) was placed in the center of the zero-background powder x-ray diffraction holder. The sample was then oriented manually to give the best reflection conditions at 2$\theta$ corresponding to (002) set of planes. After that, the x-ray diffraction pattern, containing five to seven sets of (00$l$) planes in the 10$^0$ $<2 \Theta<$ 110$^0$ range, were collected on the single crystal at room temperature using Rigaku MiniFlex powder diffractometer (Cu $K_{\alpha 1,2}$ radiation). The $c$-lattice parameters were refined by the LeBail method using Rietica software. \cite{rietica} The lattice parameters as a function of the K-concentrations (determined using $T_c$ criterion shown in the inset to Fig. \ref{c-all}) are presented in Fig. \ref{c-all}. We also plotted the data from Refs. [\onlinecite{Liu2014}] (squares) and [\onlinecite{Rotter2008a}] (triangles) for comparison. The agreement between our data and the literature is quite good. 

Thermoelectric power measurements were performed using a {\it dc}, alternating heating (two heaters and two thermometers) technique \cite{Mun2010b} by utilizing Quantum Design Physical Property Measurement System (PPMS$^{\circledR}$) to provide the temperature environment between 2 and 300 K. The samples were cleaved to get the surface of the samples as clean from the residual flux as possible. The samples were then cut and mounted directly on the surface of the SD package of the Cernox thermometers using Du-Pont 4929N silver paste to ensure thermal and electrical contact. The silver paste was allowed to cure at room temperature and 1 atm of air for about 24 h in a desiccator \cite{Hodovanets2013a}. The samples were mounted so that the thermal gradient was applied in the $ab$-plane. It should be noted that a small feature at $\sim$ 100 K in the data sets for all of the samples measured, independent of $x$-value, is probably associated with switching between two temperature ranges of the Cernox thermometer calibration curves. Two data points markedly different from the neighboring data points due to this artifact in this temperature region were removed.

\section{Results}

\begin{figure}[tbh]
\centering
\includegraphics[width=0.75\linewidth]{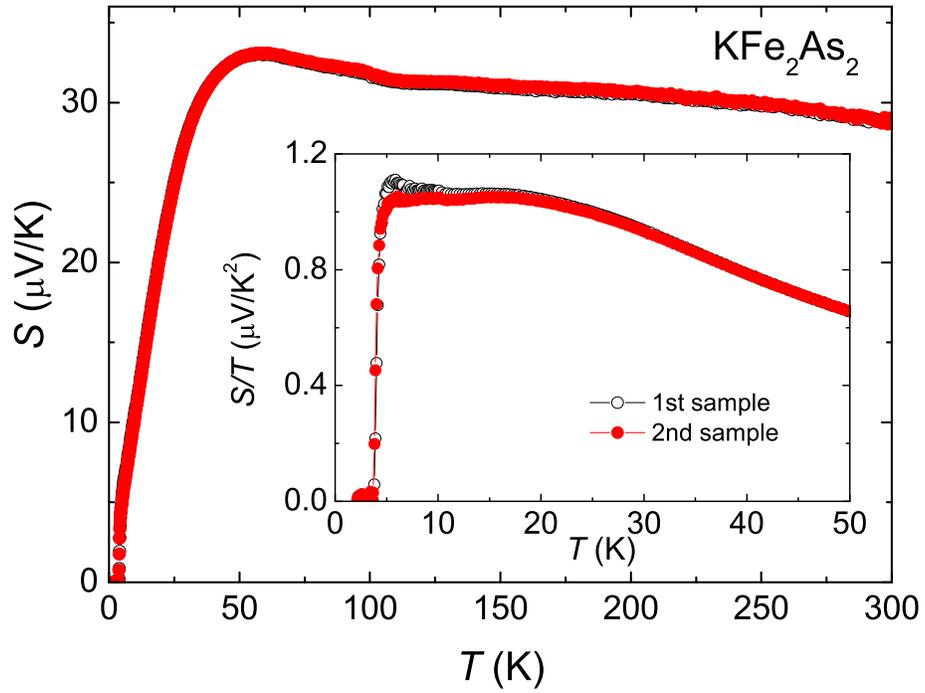}
\caption{\footnotesize (Color online) Thermoelectric power $S$ and $S/T$ (inset) data as a function of temperature of KFe$_2$As$_2$ single crystals. 
}
\label{KFe2As2}
\end{figure}

Thermoelectric power $S$ and $S/T$ data as a function of temperature for two single crystals of KFe$_2$As$_2$ are plotted in Fig. \ref{KFe2As2}. The data for these two samples are very similar except in the $S/T$ for the first sample a difference of $\sim$ 6$\%$ is observed between 5 K (above the $T_c\sim$ 4 K) and $\sim$ 20 K. It is worth pointing out that $S/T$ $\approx$ const above $T_c$ signifying Fermi-liquid-like behavior, which is consistent with the resistivity and thermal expansion measurements\cite{Hardy2013,Taufour2014,Liu2014}. 

Given that the Debye temperature for KFe$_2$As$_2$ is rather low, $\Theta_D$ = 177 K,\cite{Hardy2013} the origin of the peak at $T_{Smax} \sim$ 60 K seems unlikely to be due to phonon drag (expected at 0.1 - 0.3 $\Theta_D$\cite{Blatt1976}) but rather is probably of an electronic nature. For KFe$_2$As$_2$, a broad feature above 100 K has also been observed in the resistivity measurements, especially in the resistivity measured along the $c$-direction,\cite{Terashima2009,Liu2013} and thermal expansion measurements\cite{Hardy2013}. The origin of this feature might be associated with a coherence-incoherence crossover due to the scattering of the conduction electrons\cite{Hardy2013}. Whether the 60 K feature in the TEP is associated with this crossover or not is currently unclear.

\begin{figure}[tbh]
\centering
\includegraphics[width=0.75\linewidth]{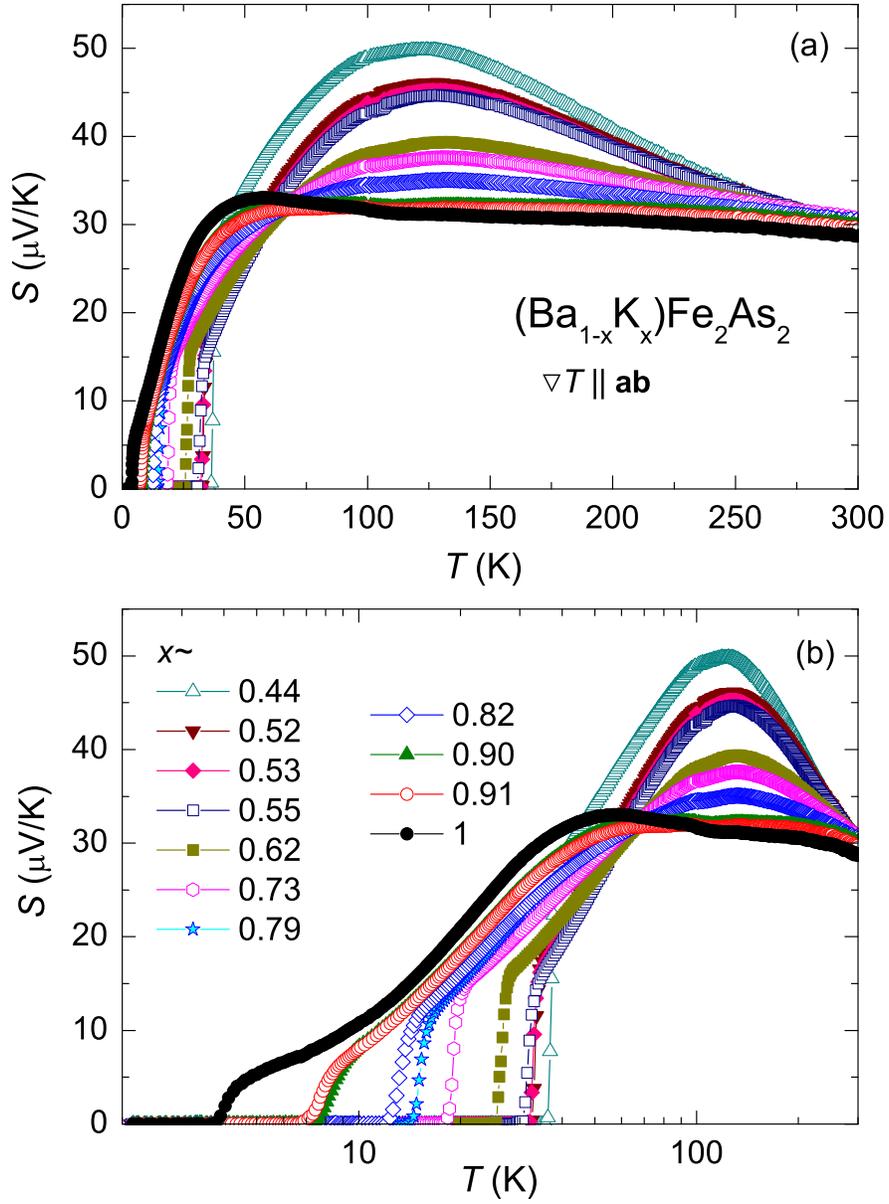}
\caption{\footnotesize (Color online) Thermoelectric power $S$ as a function of temperature of (Ba$_{1-x}$K$_x$)Fe$_2$As$_2$ (0.44$\leq x \leq$ 1) single crystals on (a) linear and (b) semi-log scales. 
}
\label{SvsT}
\end{figure}

Figures \ref{SvsT}(a) and (b) show temperature-dependent, in-plane thermoelectric power data $S$ of (Ba$_{1-x}$K$_x$)Fe$_2$As$_2$ (0.44 $\leq x \leq$ 1) single crystals on linear and semi-log scales respectively. The superconducting state for all of the samples measured can be seen as a drop to zero value in the TEP data. Above $\sim$ 100 K, as the concentration of K increases, the value of the broad maximum in the TEP is decreasing in a roughly monotonic way and the position of the maximum at $\sim$ 125 K for $x \sim$ 0.44 moves to slightly higher temperatures of $\sim$ 130 K for $x\sim$ 0.82. After $x \sim$ 0.82, for the next two potassium concentrations ($x \sim$0.9), the maximum collapses into a broad plateau followed by a maximum at $\sim$ 60 K for pure KFe$_2$As$_2$. 

The behavior of the TEP as the concentration of K is increased is more complex below $\sim$ 100 K. The change in the TEP with increasing $x$ can be better seen on a semi-log plot given in Fig. \ref{SvsT}(b). As K-concentration is increased, the value of $S$ decreases until $x\sim$ 0.55, where the value of $S$ dips and is the smallest, below $\sim$ 75 K, of all the samples measured. As the K-concentration is further increased, the low-temperature $S$ values increase again. 

It should be noted that the values of the TEP that we obtained by our measurements on single crystals are larger for heavy K-doped samples ($x\geq$ 0.8) than those reported for the polycrystalline samples\cite{Yan2010}. The difference increases as the K-concentration is increased, most likely reflecting the anisotropies in the scattering that are associated with the 2D character of most of the Fermi surfaces of the KFe$_2$As$_2$\cite{Yoshida2011} as opposed to the significantly 3D character of the Fermi surfaces of the optimally-doped Ba$_{0.6}$K$_{0.4}$Fe$_2$As$_2$\cite{Xu2011}. Indeed, the anisotropies of upper critical field $H_{c2}$ and the normal state resistivity are also the largest for KFe$_2$As$_2$. \cite{Liu2014} 

\section{Discussion}

\begin{figure}[tbh]
\centering
\includegraphics[width=0.75\linewidth]{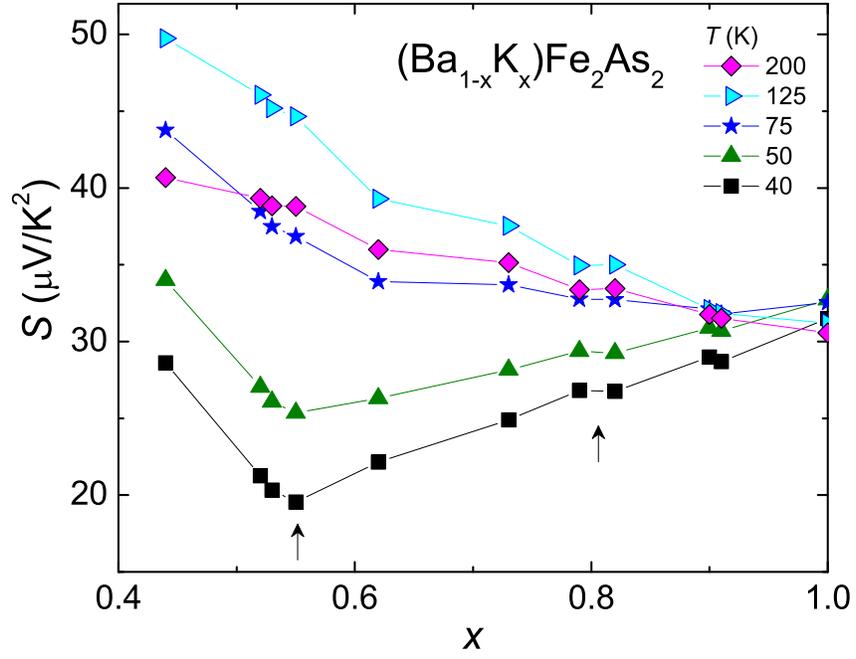}
\caption{\footnotesize (Color online) $S$ at constant temperatures of 40, 50, 75, 125, and 200 K as a function of potassium concentration of (Ba$_{1-x}$K$_x$)Fe$_2$As$_2$ (0.44$\leq x \leq$ 1) single crystals. The arrows mark a minimum at $x\sim$ 0.55 and a small feature at $x\sim$ 0.8.}
\label{Svsx}
\end{figure}

Clear features in $S\mid _{T=const}$ vs $x$ can often be associated with the Lifshitz transitions\cite{Lifshitz1960,Blanter1994,Varlamov1989} and indeed were observed in the TEP measured on Ba(Fe$_{1-x}$Co$_x$)$_2$As$_2$ single crystals \cite{Hodovanets2011,Hodovanets2013a,Mun2009}. 
To investigate the possibility of a Lifshitz transition in (Ba$_{1-x}$K$_x$)Fe$_2$As$_2$ single crystals, the data for $S$ is plotted as a function of K-concentration at constant temperatures in Fig. \ref{Svsx}. Here we used 40 K as the lowest temperature which is above $T_c$ for all $x$-values studied. Notably a minimum is clearly observable at $x\sim$ 0.55 for the two lowest temperature cuts. As the temperature is increased above 75 K, this feature becomes less pronounced and ultimately disappears. This is consistent with ARPES measurements that pointed out $x \gtrsim$ 0.6 as a concentration where the electron FS pockets at the center of $M$ point disappear.\cite{Malaeb2012} 
A multi-carrier model of the TEP could possibly explain such a change in the $S(x)$ data. However, it seems that computation of the scattering parameters and the TEP based on realistic band structure calculations that would also incorporate chemical substitution is beyond the capability of the present theory, especially for the correlated materials.
Another, more subtle, feature, a slope change, is seen at $x\sim$ 0.8. The weakness of this feature in the $S\mid _{T=const}$ vs $x$ at $x\sim$ 0.8 - 0.9 implies that there is no sharp change in the density of states at the Fermi level. The Fermi surface of (Ba,K)Fe$_2$As$_2$ is complicated. The theory of the Lifshitz transition as observed by the TEP measurements (Refs. \onlinecite{Blanter1994} and \onlinecite{Varlamov1989}) despite being extended to include the impurity effect (the dirty case) and be valid at $T >$ 0, is still only developed for the two generic types of Lifshitz transitions (void-formation and neck disruption) in a single band metal. The TEP depends not only on the derivative of the density of states at the Fermi level (the topology of the Fermi surface), but also on a type of charge carries, their mobilities and the type of scattering that they undergo, which change with temperature as well. All of this is hard to take into account when evaluating the functional dependence of the TEP near particular Lifshitz transition, especially in the multi band material. Perhaps that is why there is no general theory for the Lifshitz transitions in multiband materials, such as (Ba,K)Fe$_2$As$_2$, and no prediction on how strong or weak the feature in the TEP associated with Lifshitz transition should be. Not surprisingly, depending on the specific details of changes in the FS, the feature associated with the Lifshitz transition seen in the TEP measurements will be different. This might be especially pertinent for (Ba$_{1-x}$K$_x$)Fe$_2$As$_2$, where at $x\sim$ 0.8-0.9, hole pockets at the zone corner simultaneously transform into "four lobes"\cite{Xu2013}. 

\begin{figure}[tbh]
\centering
\includegraphics[width=0.75\linewidth]{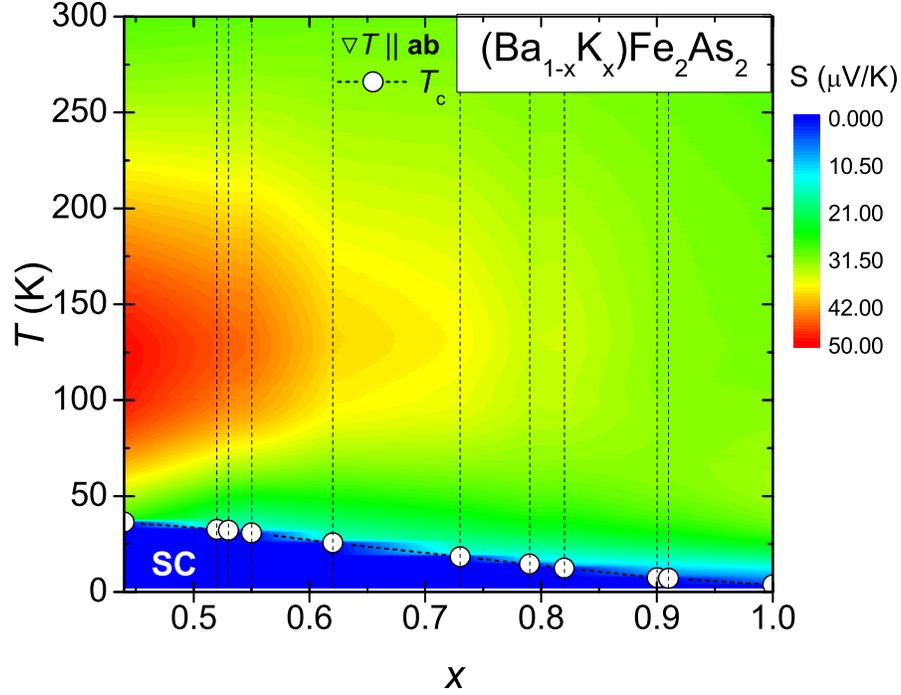}
\caption{\footnotesize (Color online) Contour plot of $S$ as a function of temperature and K-concentration. Open symbols show $T_c$-values that were determined using the offset criterion shown in the inset to Fig. \ref{c-all}. The dotted lines (as well as the two vertical axes of the edges) show where the existing data for the K-concentrations were measured. The region where the broad maximum in the $S(T)$ data evolves into a plateau is seen as an evolution from the red-yellow into green region at $x\sim$ 0.8 - 0.9.}
\label{contoura}
\end{figure}

To visualize the change in $S$ at $x\sim$ 0.8-0.9 more clearly, we can look at more than just isothermal $S(x)$ cuts; we can create a broader overview by interpolating or extrapolating our $S(x,T)$ data and create a 2D contour plot,\cite{Origin} Fig. \ref{contoura}, where $S$ is a function of temperature and K-concentration. 
The $T_c - x$ phase diagram and the dotted lines indicating the concentrations of the measured samples were also added to the plot. It is notable that the red-yellow hight $S$ region evolves into a green region in $S(x,T)$ at roughly $x\sim$ 0.8 - 0.9, and indicates, as we discussed above, that the broad maximum in $S(T)$ evolves into a plateau.

\begin{figure}[tbh]
\centering
\includegraphics[width=0.75\linewidth]{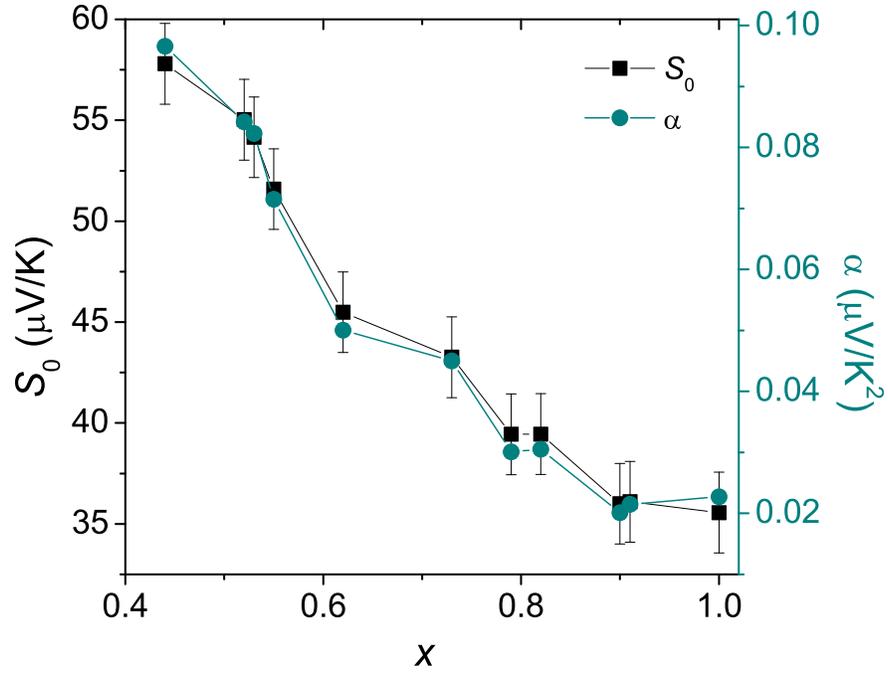}
\caption{\footnotesize (Color online) $S_0$ and $\alpha$ as a function of K-concentration. $S_0$ and $\alpha$ are the results of the linear fit, $S$ = $S_0$ - $\alpha$ $T$, to $S(T)$ over 250 K$ <T<$ 300 K.}
\label{Soslope}
\end{figure}

\begin{figure}[tbh]
\centering
\includegraphics[width=0.75\linewidth]{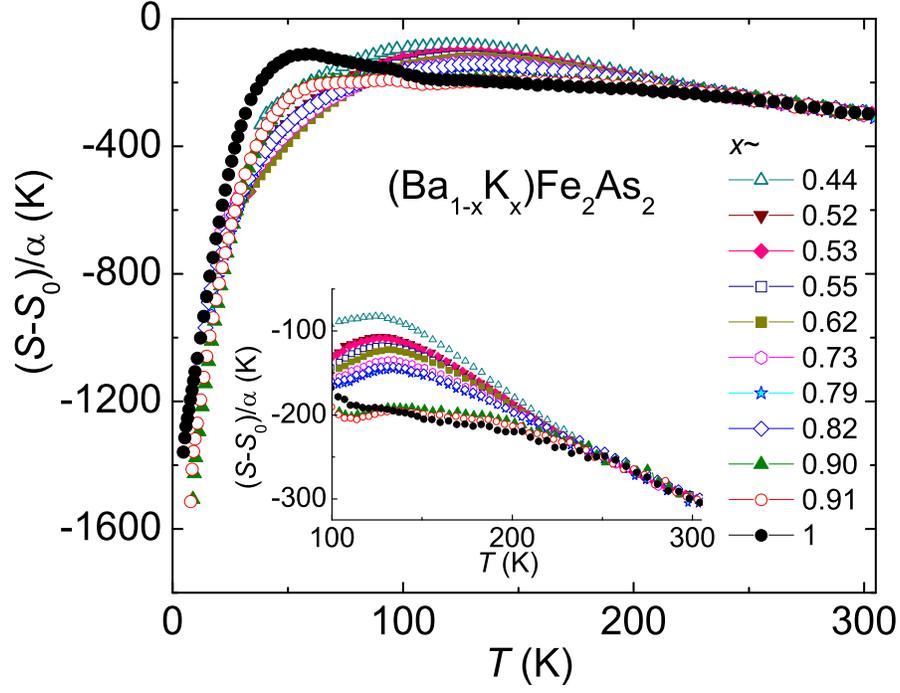}
\caption{\footnotesize (Color online) Thermoelectric power $S$ as a function of temperature of (Ba$_{1-x}$K$_x$)Fe$_2$As$_2$ single crystals scaled based on the linear fit, $S$ = $S_0$ - $\alpha T$, over 250 K$ <T<$ 300 K. The inset shows the 100 K $\leq T \leq$ 300 K data on an enlarged scale. For clarity of the data, every sixth data point is plotted and the superconducting region is excluded.}
\label{scaling}
\end{figure}

The TEP data above $T > \Theta_D$ can be analysed in a yet another way: for example, for high-$T_c$ cuprates\cite{Mandal1996,Kim2004}, the scaling of the $S(T)$ data based on the linear fit of the high temperature TEP data was done. In order to scale the $S(T)$ data in a manner similar to high-$T_c$ cuprates, we performed a linear fit, $S$ = $S_0$ - $\alpha T$, to the $S$ vs $T$ data over the temperature range 250 K $<T<$ 300 K. Figure \ref{Soslope} shows $S_0$ and $\alpha$ as a function of K-concentration. The evolutions of $S_0$ and the slope $\alpha$ with the increase of K content are similar and rather monotonic. However, two breaks in the slope at $x\sim$ 0.6 and $x\sim$ 0.9 can be seen and are at similar $x$-values discussed above. The scaling of the data to the high-temperature slope, $\alpha$, is shown in Fig. \ref{scaling}. It worth noting that below $\sim$ 250 K, the data are split into two manifolds at $x\sim$ 0.8 - 0.9, again emphasizing some change in the TEP associated with this doping level. 
This split into two manifolds further supports the conclusion that $x \sim$ 0.8-0.9 is a critical concentration range. We think that these observations indicate a change in the Fermi surface topology and are consistent with or related to the BNC scaling deviation\cite{Budko2013} that was observed in the specific heat measurements, prediction of the Lifshitz transition by the band structure calculations\cite{Khan2014}, ARPES measurements\cite{Xu2013, Ota2014} and with the change in the superconducting gap symmetry as observed by thermal conductivity and penetration depth\cite{Watanabe2014} measurements. 

\section{Conclusion}

We measured thermoelectric power on (Ba$_{1-x}$K$_x$)Fe$_2$As$_2$ single crystals in the over-doped side of the $T-x$ phase diagram, 0.44 $\leq x \leq$ 1. We observe a minimum in the $S\mid _{T=const}$ vs $x$ at $x \sim$ 0.55 that can be associated with the change in the topology of the Fermi surface, significant changes in the electronic structure, correlations, and/or scattering. This feature is clearly observable below $\sim$ 75 K and a higher temperature smears it out. Also, $S$ shows a change in the $x \sim$ 0.8 - 0.9 range, where a broad maximum in $S(T)$ evolves into a plateau, seen better in the contour plot of $S(x,T)$ as an evolution of the red-yellow high $S$ region into a green region. In addition, the scaled thermoelectric power data are separated into two manifolds at high temperature at $x\sim$ 0.8 - 0.9 range as well. The features seen in the TEP measurement are consistent with the Lifshitz transitions observed by ARPES measurements,$i.e$, the Lifshitz transition at $x\sim$ 0.55 may be associated with the shift of the electron pockets at the $M$ point above the Fermi level\cite{Malaeb2012} and the Lifshitz transition at $x\sim$ 0.8 - 0.9 may be associated with the transformation of the hole pockets near the $M$ point into "four blades"\cite{Xu2013} or other significant changes of the electronic structure or correlations might occur.

\section{ACKNOWLEDGEMENTS}

The authors would like to thank R. Prozorov and M. A. Tanatar for bringing our attention to the possibility of these measurements; B. Coles for pointing out the versatility of this measurement technique;  H. Kim, V. Taufour and M. Sailer for useful discussions. This work was supported by the U.S. Department of Energy (DOE), Office of Science, Basic Energy Sciences, Materials Science and Engineering Division. The research was performed at the Ames Laboratory, which is operated for the U.S. DOE by Iowa State University under contract $\#$ DE-AC02-07CH11358. E. Mun was supported by the AFOSR-MURI grant No. FA9550-09-1-0603.

%

\end{document}